\documentclass[prl,floats,showpacs,aps,twocolumn,floatfix,amsmath]{revtex4}
\usepackage{graphicx}
\begin{document}
\title{Fractal Weyl laws for amplified states in ${\cal PT}$-symmetric resonators}

\author{Christopher Birchall}
\author{Henning Schomerus}
\affiliation{Department of Physics, Lancaster University, Lancaster,  LA1 4YB, United Kingdom}

\date{\today}

\begin{abstract}
We find that in nonhermitian ${\cal PT}$-symmetric systems (as realized
in resonators with balanced absorption and amplification), a mechanism
based on quantum-to-classical correspondence reduces the occurrence of
strongly amplified states. The reduction arises from semiclassically
emerging hierarchical phase-space structures that are associated with the
coupling of the amplifying and absorbing regions (forward and
backward-trapped sets and their complements), and amounts to a
generalization of the fractal Weyl law, earlier proposed for
ballistically open systems. In the context of the recently introduced
class of ${\cal PT}$-symmetric laser-absorbers, this phenomenon reduces
the number of states participating in the mode competition.
\end{abstract}

\maketitle

Nonhermitian systems which can still possess a real spectrum have been
realized  very recently in optical \cite{experiments1a,experiments1b}
electronic \cite{experiments2} and microwave \cite{passive} settings.
These experiments utilize the concept of $\mathcal{PT}$ symmetry
\cite{bender,makris}---when an amplifying system is coupled symmetrically
to an absorbing system, this can result in a precise balance of loss and
gain for individual modes, which then have a real energy (the time
reversal operation $\mathcal{T}$ converts loss into gain and vice versa,
while the parity operation ${\cal P}$ interchanges the amplifying and
absorbing parts of the system). For small absorption and amplification
rates $\mu$, this balance is robust, but if these rates are too large,
pairs of real eigenenergies bifurcate into complex-conjugate pairs. This
phenomenon is known as spontaneous $\mathcal{PT}$-symmetry breaking and
leads to a range of remarkable switching effects
\cite{experiments1a,experiments1b,makris,pteffects1,pteffects2,pteffects3}.
In particular, $\mathcal{PT}$-symmetric optical resonators have been
predicted to form a new class of lasers \cite{hs,longhi,stone}, which
show an additional spectral peculiarity: lasing modes are degenerate with
perfectly absorbing modes \cite{longhi,stone,stone2}. In the setting of
these lasers, the focus shifts to the most strongly amplified modes, as
these are best placed to overcome external losses, win the mode
competition, and thus determine the laser threshold and radiation
characteristics.

Here we report that the most strongly amplified modes in  these
$\mathcal{PT}$-symmetric resonators show the hallmarks of yet another
distinctive spectral feature---a \emph{fractal Weyl law}, by which the
number of these modes is systematically reduced when the resonator
dimensions are scaled up to become much larger than the wave length (the
semiclassical limit, corresponding to an effective Planck's constant
$h\ll 1$). Fractal Weyl laws have been originally established for passive
leaky systems with chaotic underlying classical dynamics, where
long-living states are supported by a fractal repeller---the number of
these states then scales as $h^{d_H}$, where $d_H<d$ is the fractal
dimension of the repeller in the $d$-dimensional phase space
\cite{weyl1,weyl2,weyl3,weyl4,weyl5}. In order to uncover the analogous
effects in the $\mathcal{PT}$-symmetric setting, we establish a
quantum-to-classical correspondence \cite{weyl5,husimischur,graefe} of
various components of the spectrum to specific regions in the classical
phase space, with phase-space volumes reflecting the proportions of these
components in the spectrum. We find that the strongly amplified states
are supported by a hierarchical structure associated with the coupling of
the amplifying and absorbing regions (the backward-trapped set in the
amplifying parts of the system, which forms part of the classical
repeller). The increasing phase-space resolution in the semiclassical
limit reveals the sparse fractal nature of the backward-trapped set and
thus results in the reduction of the number of strongly amplified modes
mentioned above.

\begin{figure}[b]
\includegraphics[width=.8\columnwidth]{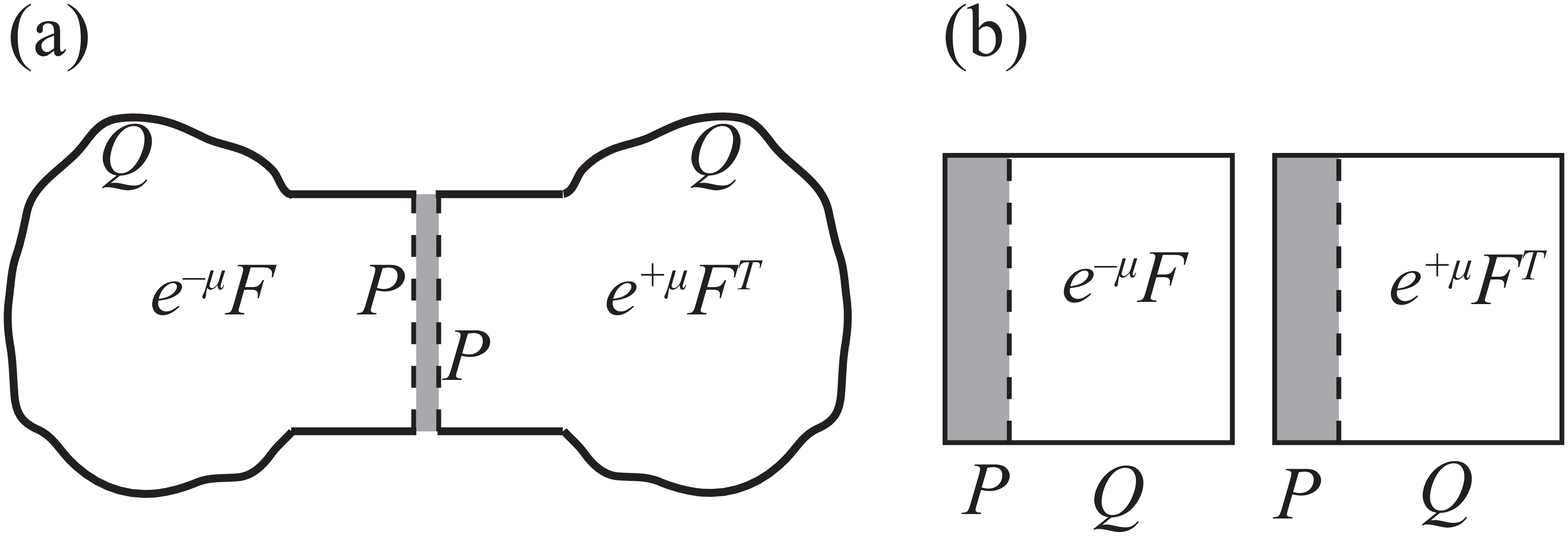}
\caption{\label{fig1}
(a) Sketch of a $\mathcal{PT}$-symmetric resonator, consisting of an absorbing resonator (left) coupled to an amplifying resonator (right). The unitary operator $F$ describes the internal dynamics in each resonator, the factors $e^{\pm\mu}$ model amplification and absorption, and the projectors $P$ (onto the interface) and $Q$ (onto the walls) describe the coupling,
altogether resulting in the quantum map \eqref{eq:map}.
(b) The Hilbert space (and the classical phase space) divides into an absorbing and an amplifying part, coupled in a subspace (shaded gray).
}
\end{figure}

\emph{Model for nonhermitian $\mathcal{PT}$-symmetric resonator
dynamics.---} A model of the dynamics of waves in a resonator which
displays the required $\mathcal{PT}$ symmetry can be set up as follows
(see Fig.~\ref{fig1}). We divide the Hilbert space into two subspaces
$L$, $R$, each of dimension $M$, with the first subspace representing the
absorbing region in a `left' resonator and the second subspace
representing the amplifying region in a `right' resonator. In each
resonator, the ballistic chaotic dynamics over a finite time interval
$\Delta t \equiv 1$ is described by an $M\times M$-dimensional unitary
time-evolution operator $F_L=F$ and $F_R=[F^{-1}]^*=F^T$, respectively,
where the stated relation between $F_L$ and $F_R$ ensures $\mathcal{PT}$
symmetry \cite{hs,stone}. In each time step, the wave amplitude in the
absorbing region is reduced by a factor $\exp(-\mu)$, while in the
amplifying region it is enhanced by $\exp(\mu)$. This breaks the
unitarity of the time evolution, but because of the matching rates again
respects $\mathcal{PT}$ symmetry. Finally, the ballistic coupling between
the subspaces is described by a coupling matrix
\begin{equation}
C=\left(\begin{array}{cc} Q &   -iP \\  -iP & Q \\ \end{array}  \right),\quad
P=\mathrm{diag}(\underbrace{1,\ldots,1}_N,\underbrace{0,\ldots,0}_{M-N}).
\label{eq:c}
\end{equation}
where  $P$ projects onto the interface between the two resonators (with
$N$ channels connecting the resonators), while $Q\equiv \openone_M-P$
represents the projector onto the walls of each resonator.

With these specifications, the time evolution of the composed system can
be written as
\begin{equation}
{\cal F}=
\sqrt{C}\left(\begin{array}{cc} e^{-\mu}F & 0 \\ 0 &  e^{\mu}F^T \\ \end{array}  \right)\sqrt{C},
\label{eq:map}
\end{equation}
where we symmetrized the coupling by means of the unitary matrix
$\sqrt{C}=\left(\begin{array}{cc} 2^{-1/2}P+Q & -i2^{-1/2}P \\
-i2^{-1/2}P  &  2^{-1/2}P+Q \\ \end{array}  \right)$ which squares to
$C$. The operator ${\cal F}$ acts on $2M$-dimensional vectors
$\psi=\binom{\psi_L}{\psi_R}$, with the $M$ entries in $\psi_L$ giving
the wave amplitude in the absorbing subsystem, and the remaining $M$
entries in $\psi_R$ giving the wave amplitude in the amplifying
subsystem.

The spectrum of the system is now obtained from the eigenvalue equation
\begin{equation}\label{eq:lambda}
{\cal F}\psi_n=\lambda_n\psi_n,\quad \lambda_n=\exp(-iE_n),
\end{equation}
where $E_n=\varepsilon_n-i\Gamma_n/2$ are quasienergies. The
$\mathcal{PT}$ symmetry of the quantum map \eqref{eq:map} is manifest by
the relation
\begin{equation}{\cal F}=\mathcal{P}[{\cal F}^{-1}]^*\mathcal{P},
\end{equation}
where the parity operator is of the explicit form
$\mathcal{P}=\left(\begin{array}{cc} 0 &   \openone_M \\  \openone_M & 0
\\ \end{array}  \right)=\openone_M\otimes\sigma_x$ (thus, $\mathcal{P}$
interchanges the amplitudes of the subsystems). The spectral properties
associated with $\mathcal{PT}$ symmetry are now embodied in the
characteristic polynomial $s(\lambda)=\rm{det}\,({\cal F} -\lambda)$, of
degree $2M$. Due to  $\mathcal{PT}$ symmetry, this polynomial exhibits
the mathematical property of \emph{self-inversiveness}
\cite{selfinverse1,selfinverse2}:
\begin{equation}
s(1/\lambda^*)=[\lambda^{-2M}s(\lambda)]^*s(0),
\end{equation}
where $s(0)=\mathrm{det}\,\mathcal{F}= (\mathrm{det}\,F)^2$ is a
unimodular complex number. The eigenvalues are the roots of the secular
equation $s(\lambda)=0$. For each eigenvalue $\lambda_n$, we are thus
guaranteed to find the eigenvalue $\lambda_{\bar n}=[\lambda_n^{-1}]^*$.
It is possible that $n=\bar n$; then $|\lambda_n|=1$, which means that
the quasienergy $E_n=i\ln \lambda_n$ is real. For $n\neq \bar n$, we
obtain complex quasienergies, which thus still appear in pairs ($E_{\bar
n}=E_n^*)$. If $\mu=0$, $\mathcal{F}$ is unitary, and all quasienergies
are real. In the semiclassical limit of large $M=h^{-1}$ and fixed
classical coupling strength $N/M$, and with $\mu$ increasing from zero,
one expects a quasi-monotonous increase of the fraction of complex
energies.

\emph{Dynamical signatures of spontaneous $\mathcal{PT}$-symmetry
breaking.---} We now investigate how the transition from a real to a
complex spectrum depends on the dynamics of the system. In the resonator
setting, nonintegrable dynamics arise from resonator walls of a generic
shape. In the quantum map \eqref{eq:map}, this can be modeled by choosing
a suitable operator $F$ for the internal dynamics. In an optical setting,
the passive system is time-reversal invariant, which furthermore dictates
$F=F^T$. We implement these features via a paradigm of wave dynamics with
a nonintegrable classical limit, the kicked rotator \cite{kr,model}. The
time-evolution operator of this system is given by
\begin{equation}
F_{mm'}=(iM)^{-1/2}e^{\frac{i\pi}{M}(m-m')^2-\frac{iMk}{4\pi}\left(\cos\frac{2\pi m}{M}+\cos\frac{2\pi m'}{M}\right)},
\label{eq:rotor}
\end{equation}
where the kicking strength $k$ controls the  dynamics from classically
integrable ($k=0$) to globally chaotic ($k\gtrsim 7$); the classical map
is $q'=q+p+(k/4\pi)\sin(2\pi q) \pmod 1$, $p'=p+(k/4\pi)[\sin(2\pi
q)+\sin(2\pi q')] \pmod 1$. We focus on the chaotic cases with $k=8$, fix
the coupling of the absorbing and amplifying regions by setting the
inverse average dwell time in the amplifying or absorbing regions (the
Thouless energy) to $E_T=N/M=1/5$, and analyze the spectrum (obtained by
diagonalizing ${\cal F}$) as a function of the amplification rate $\mu$
and the system size $M$.

\begin{figure}
\includegraphics[width=\columnwidth]{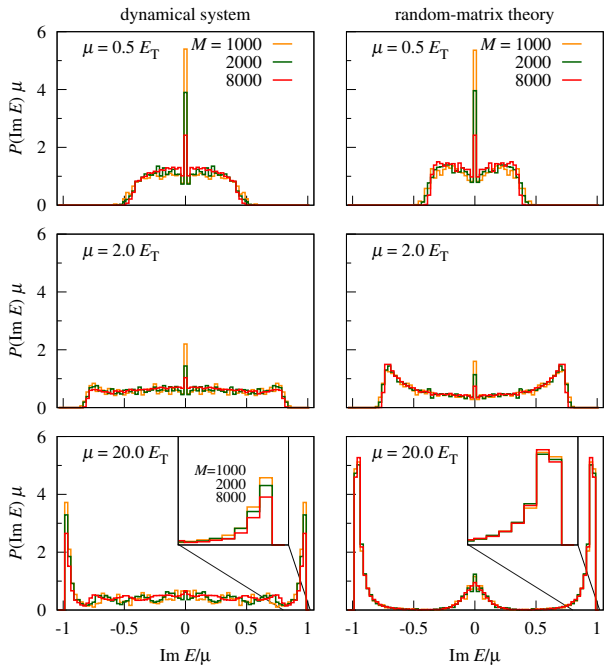}
\caption{\label{fig2} (Color online)
Distributions of ${\rm Im}\,E$ in a $\mathcal{PT}$-symmetric resonator
with Thouless energy $E_T\equiv N/M=1/5$ for various amplification rates
$\mu$ and system sizes $M$, numerically obtained from Eqs.\
\eqref{eq:map}, \eqref{eq:lambda} with internal dynamics modeled via the
kicked rotator \eqref{eq:rotor} with $k=8$ (left) and via random-matrix
theory (right). Significant differences occur when  $\mu$ exceeds $E_T$.
For the kicked rotator, the proportion of strongly amplified states
decreases systematically with increasing $M$, as highlighted in the inset
of the bottom panel. }
\end{figure}

\begin{figure}
\includegraphics[width=\columnwidth]{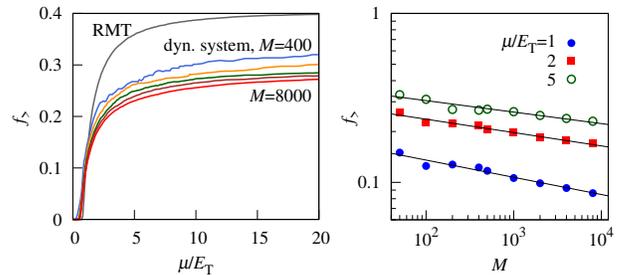}
\caption{\label{fig3} (Color online)
Left panel: Fraction $f_>(\mu)$ of strongly amplified states with ${\rm Im}\,E>\mu/2$, as a function of $\mu$.
The random-matrix curve (RMT) is evaluated at $M=8000$, while the curves for the dynamical system (specified in Fig.~\ref{fig2})
correspond to $M=400$, $1000$, $2000$, $4000$ and $8000$.
The double-logarithmic plot in the right panel demonstrates power-law
scaling $f_>\propto M^{-a}$ at fixed $\mu=E_T$ ($a=0.102(9)$),  $2\,E_T$ ($a=0.079(6)$), and
$5\,E_T$ ($a=0.069(8)$). }
\end{figure}

We start with the distribution of decay rates, encoded in ${\rm Im}\,E$,
where large positive  values indicate strong amplification and large
negative values indicate strong decay. The left panels in Fig.\
\ref{fig2} show histograms of this quantity for three representative
values of $\mu$. For comparison, the right panels show results from
random-matrix theory, with $F$ taken from the circular orthogonal
ensemble \cite{mehta}. All histograms are symmetric because complex
energies appear in conjugate pairs, as imposed by $\mathcal{PT}$
symmetry. Furthermore, the histograms display a sharp peak at ${\rm
Im}\,E=0$, which arises from the states with real energies.  This peak
decreases with increasing systems size $M$, indicating that the
proportion of such states vanishes in the semiclassical limit. As regards
to this peak, there is good correspondence between the dynamical system
and random-matrix theory, which predicts that the spectrum essentially
turns complex at $\mu_c=\sqrt{N}/M\ll E_T$ \cite{hsrmt}.

For $\mu=0.5 E_T$ (top panel), this agreement between both models also
extends to finite  values of ${\rm Im}\, E$, even though some difference
are noticeable. The differences become more marked as $\mu$ exceeds $E_T$
(middle and bottom panels)---in particular, in comparison to
random-matrix theory, the dynamical system possesses a significantly
reduced number of strongly amplified states, with large ${\rm Im}\, E$.
The insets in the bottom panel focus on the peak associated with these
states. In the dynamical system, their proportion decreases
systematically as $M$ increases, while in random-matrix theory their
proportion does not change.

In order to quantify this systematic reduction we consider the fraction
of states with  ${\rm Im}\, E>\mu/2$, denoted as $f_{>}$ and plotted as a
function of $\mu$ in the left panel of Fig.\ \ref{fig3}. In random-matrix
theory, this curve becomes independent of $M$ when $M$ is large. For the
dynamical system, however, $f_>(\mu)$ drops as $M$ increases, even at the
largest computationally accessible system sizes ($M=8\,000$,
corresponding to a matrix dimension $16\,000$ for ${\cal F}$). As shown
in the right panel this drop follows a power-law $f_>(\mu)\propto
M^{-a}$, with $a=0.102(9)$ ($\mu=E_T$), $a=0.079(6)$ ($\mu=2E_T$), and
$a=0.069(8)$ ($\mu=2E_T$). This data confirms that deviations from
random-matrix theory set in as $\mu$ exceeds $E_T$. That the deviations
increase with $M$ points to a semiclassical origin, which we identify
next.

\begin{figure}
\includegraphics[width=\columnwidth]{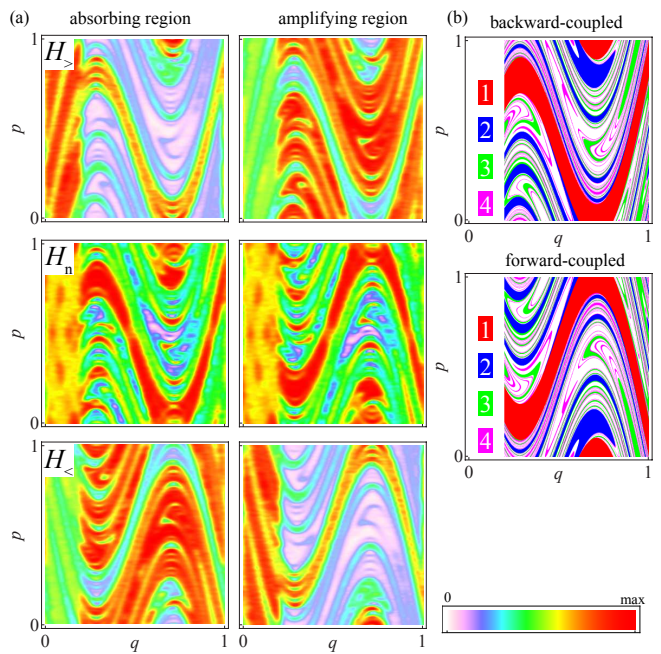}
\caption{\label{fig4} (Color online)
(a) Husimi phase-space distributions of the support of strongly amplified
states (${\rm Im}\, E>\mu/2$, $H_>$), `neutral' states ($|{\rm Im}\,
E|<\mu/2$, $H_n$), and strongly decaying states (${\rm Im}\, E<-\mu/2$,
$H_<$), obtained for the kicked-rotator model of a
$\mathcal{PT}$-symmetric resonator with $k=8$, $M=1000$, $E_T=1/5$,
$\mu=2\,E_T$. (b) Classical phase-space regions which are coupled to the
interface by $1-4$ steps, backwards in time (top panel) or forwards in
time (bottom panel). }
\end{figure}

Following previous investigations of general non-hermitian systems that
addressed the issue of state nonorthogonality  \cite{weyl5,husimischur},
we identify the semiclassical support of the eigenstates in different
components of the spectrum by constructing an orthonormal basis $\phi_n$
of the subspace spanned by the states, and then summing the Husimi
phase-space representations $H(q,p)$ of the basis states, defined by the
squared overlap $|\langle q,p|\phi_n\rangle|^2$ with minimal-uncertainty
wavepackets centered at phase space position $(q,p)$.
Figure~\ref{fig4}(a) shows  these distributions  for $\mu=2\,E_T$ and
$M=1000$. The top panel depicts the support $H_>$ of the states with
${\rm Im}\, E>\mu/2$, which contribute to the quantity $f_>(\mu)$
discussed above.  The middle panel shows the support $H_n$ of `neutral'
states with  $|{\rm Im}\, E|<\mu/2$, while the bottom panel shows the
support  $H_<$  of quickly decaying states  with ${\rm Im}\, E<-\mu/2$.
The distributions $H_>$ and $H_<$ are related by $\mathcal{PT}$ symmetry,
which interchanges the absorbing and amplifying regions and maps the
momentum  $p\to -p$; $H_n$ is invariant under these operations.

The systematic  structures observed in these distributions have a
specific classical dynamical interpretation, which follows from the plots
shown in panel (b) of Fig.\  \ref{fig4}. The top panel in (b) shows
\emph{backward-coupled regions} which are populated from the interface
after $1 - 4$ times steps of the classical dynamics. Including higher
iterations, this hierarchical structure defines the complement of the
\emph{backward-trapped set}, a fractal which contains all points whose
backwards-in-time dynamics does not meet the interface. The lower panel
in (b) shows the analogous \emph{forward-coupled regions}, which
propagate towards the interface and define the complement of the
\emph{forward-trapped set} (the time reverse of the backward-trapped
set).

Coming back to panel (a), we now recognize that in the amplifying part of
phase space, the strongly amplified states mainly populate the
backward-trapped set, and thus  concentrated in regions of maximal dwell
time. In the absorbing part of the system, these states
%have a much
%reduced weight, $|\psi_L|^2/|\psi_R|^2\sim [e^{2(\mu-{\rm
%Im}\,E)}-1]/[1-e^{-2(\mu+{\rm Im}\,E)}]$,
%and
 are supported by the
backward-coupled regions, which signifies minimal dwell time in this part
of the system. A power-law suppression $\propto M^{-a}$ of the number of
these states now follows from the fractal nature of the backward-trapped
set, which is resolved in more detail in the semiclassical limit
(according to the shrinking size $1/M$ of a Planck cell in phase space).
The relevant fractal repeller dimension can then be estimated from the
data in Fig.~\ref{fig3} as $d_H=2-a$ (the precise definition of this
dimension in the context of fractal Weyl laws is an open problem). For
larger values of $\mu$, $f_>$ picks up larger parts of the bulk of the
distribution $P(\mathrm{Im}\,E)$, thus reducing the slope and
overestimating $d_H$. We note that there are quantum fluctuations as a
function of $M$, and classical finite-size effects as the dwell time
$E_T^{-1}=5$ is not asymptotically large. For the strongly decaying
states, the same argumentation carries over to the forward-trapped set.
Furthermore, the neutral states are supported by the backward-coupled
regions in the amplifying part, and the forward-coupled regions in the
absorbing parts, thus resulting in an (approximate) balance of
amplification and absorption which becomes exact for states with ${\rm
Im}\, E=0$.

In conclusion, the spectral properties of $\mathcal{PT}$-symmetric
systems with complex wave dynamics show clear signatures of the
underlying classical dynamics, and in particular, the details of the
coupling between amplifying and absorbing regions. Strongly amplified
states are supported by the backward-trapped set in the amplifying
region, while states with real or almost real energies are supported by
regions that are well connected to both the amplifying and the absorbing
parts of the system. The fractal nature of the backward-trapped set
results in a systematic reduction of strongly amplified states, which
becomes more marked for large system sizes, and follows the
characteristic power-law dependence of a fractal Weyl law (earlier
predicted only for leaky systems with ballistic openings).

The experimental observation of the ordinary fractal Weyl law has proven
a challenge as it addresses decaying quasi-bound states. In contrast, the
fractal Weyl law uncovered here applies to amplified and possibly lasing
states. This phenomenon should generally affect the properties of the
discussed class of $\mathcal{PT}$-symmetric laser-absorbers,  where the
strongly amplified states dominate the mode competition. However,
conceptually our findings also carry over to other classes of
nonhermitian dynamical systems, including passive realizations of
$\mathcal{PT}$ symmetry \cite{experiments1a,passive}, as well as systems
which combine amplification and absorption in a non-symmetrical fashion
\cite{ampabs}. The main requirements are multiple scattering on long time
scales and ballistic wave propagation on short time scales, including a
ballistic coupling between the regions of different amplification or
absorption rate; the most amplified (or least decaying) states are then
semiclassically supported by the backward-trapped set in the most
amplifying (or least absorbing) regions.
%This also applies to systems
%with a mixed phase space, where the support may include classical islands
%of stability, and thus define a fat fractal.


\vspace*{-.5cm}
\begin{thebibliography}{99}
\vspace*{-.5cm}

\bibitem{experiments1a}
A. Guo, G. J. Salamo, D. Duchesne, R. Morandotti, M. Volatier-Ravat, V. Aimez, G. A. Siviloglou, and D. N. Christodoulides,
Phys. Rev. Lett. \textbf{103}, 093902 (2009).
\bibitem{experiments1b}
 C. E. R{\"u}ter, K. G. Makris,
R. El-Ganainy, D. N. Christodoulides, M. Segev, and D. Kip,
Nature Phys. \textbf{6}, 192 (2010).

\bibitem{experiments2}
%Experimental study of active LRC circuits with PT symmetries
J. Schindler, A. Li, M. C. Zheng, F. M. Ellis, and T. Kottos,
Phys. Rev. A \textbf{84}, 040101 (2011).


\bibitem{passive} S. Bittner, B. Dietz, U. G{\"u}nther, H. L. Harney,
    M. Miski-Oglu, A. Richter, and F. Sch{\"a}fer, Phys. Rev. Lett.
    \textbf{108}, 024101 (2012).


\bibitem{bender}
C. M. Bender and S. Boettcher, Phys. Rev. Lett. \textbf{80}, 5243 (1998).

\bibitem{makris}
R. El-Ganainy, K. G. Makris, D. N. Christodoulides, and
Z. H. Musslimani, Opt. Lett. \textbf{32}, 2632 (2007); K. G. Makris,
R. El-Ganainy, D. N. Christodoulides, and Z. H. Musslimani,
Phys. Rev. Lett. \textbf{100}, 103904 (2008);
Z. H. Musslimani, K. G. Makris, R. El-Ganainy, and D. N.
Christodoulides, \emph{ibid.} \textbf{100}, 030402 (2008).


\bibitem{pteffects1}
S. Longhi, Phys. Rev. Lett. \textbf{103}, 123601 (2009).
\bibitem{pteffects2}
M. C. Zheng, D. N. Christodoulides, R. Fleischmann, and
T. Kottos, Phys. Rev. A \textbf{82}, 010103 (2010).
\bibitem{pteffects3}
A. A. Sukhorukov, Z. Xu, and Y. S. Kivshar, Phys. Rev. A \textbf{82},
043818 (2010).

\bibitem{hs} H. Schomerus, Phys. Rev. Lett. \textbf{104}, 233601 (2010);
G. Yoo, H.-S. Sim, and H. Schomerus, Phys. Rev. A \textbf{84}, 063833 (2011).

\bibitem{longhi} S. Longhi, Phys. Rev. A \textbf{82}, 031801(R) (2010).


\bibitem{stone}
Y. D. Chong, L. Ge, and A. D. Stone, Phys. Rev. Lett. \textbf{106}, 093902 (2011);
L. Ge, Y. D. Chong, and A. D. Stone, Phys. Rev. A \textbf{85}, 023802 (2012).

\bibitem{stone2}
Y. D. Chong, L. Ge, H. Cao, and A. D. Stone,  Phys. Rev. Lett. \textbf{105}, 053901 (2010);
W. Wan, Y. Chong, L. Ge, H.  Noh, A. D. Stone, and H. Cao, Science \textbf{331}, 889 (2011).

\bibitem{weyl1}
J. Sj{\"o}strand, Duke Math. J. \textbf{60}, 1 (1990).

\bibitem{weyl2}
 M. Zworski, Not.
Am. Math. Soc. \textbf{46}, 319 (1999);
W. T. Lu, S. Sridhar, and M. Zworski, Phys. Rev. Lett. \textbf{91}, 154101 (2003);
S. Nonnenmacher and M. Zworski, J. Phys. A \textbf{38}, 10683 (2005).
\bibitem{weyl3}
H. Schomerus and J. Tworzyd{\l}o, Phys. Rev. Lett. \textbf{93}, 154102 (2004).

\bibitem{weyl4}
J. P. Keating, M. Novaes, S. D. Prado, and M. Sieber, Phys. Rev. Lett. \textbf{97}, 150406 (2006).
\bibitem{weyl5}
L. Ermann, G. G. Carlo, and M. Saraceno, Phys. Rev. Lett. \textbf{103}, 054102 (2009).

\bibitem{husimischur}
%Fractal Weyl laws for quantum decay in dynamical systems with a mixed phase space
M. Kopp and H. Schomerus,
Phys. Rev. E \textbf{81}, 026208 (2010).

\bibitem{graefe}
E.-M. Graefe, M. H\"oning, and H. J. Korsch, J. Phys. A: Math. Theor. \textbf{43}, 075306 (2010);
E.-M. Graefe and R. Schubert, Phys. Rev. A \textbf{83}, 060101(R) (2011).

\bibitem{selfinverse1}
E. Bogomolny, O. Bohigas, and P. Leboeuf, Phys. Rev.
Lett. \textbf{68}, 2726 (1992).
\bibitem{selfinverse2}
 F. Haake, M. Ku\'s, H.-J. Sommers, H. Schomerus, and
K. \.Zyczkowski, J. Phys. A \textbf{29}, 3641 (1996).


\bibitem{kr}
F. M. Izrailev, Phys. Rep. \textbf{196}, 299 (1990).

\bibitem{model}
For a related  waveguide model
see C. T. West, T. Kottos, and T. Prosen, Phys. Rev. Lett. \textbf{104}, 054102 (2010).

\bibitem{mehta} M. L. Mehta, \emph{Random Matrices}, 3rd ed. (Elsevier, New York, 2004).

\bibitem{hsrmt}
%Universal routes to spontaneous PT-symmetry breaking in non-Hermitian quantum systems
H. Schomerus, Phys. Rev. A \textbf{83}, 030101(R) (2011).




%\bibitem{orthnote}
%Note that the nonhermitian violation of orthogonality results in a finite overlap of the semiclassical support of the various groups of states.



\bibitem{ampabs}
L. Ge, Y. D. Chong, S. Rotter, H. E. T{\"u}reci, and A. D. Stone,
Phys. Rev. A \textbf{84}, 023820 (2011).

%\bibitem{nonnenmacherabsorb} S. Nonnenmacher and E. Schenck, Phys. Rev. E
%    \textbf{78}, 045202(R) (2008).
\end{thebibliography}
\end{document}